\documentclass[pra,preprint,showpacs]{revtex4}
\usepackage[centertags]{amsmath}
\usepackage{amsfonts}
\usepackage{amssymb}
\usepackage{amsthm} 
\usepackage{newlfont}
\usepackage{epsfig}
\usepackage{amscd}
\usepackage{graphicx}
\usepackage{epsfig}
\usepackage{footnote}
\usepackage{lipsum}
\usepackage{color}
\usepackage{xcolor}
\usepackage{graphicx}
\usepackage{multirow}
\usepackage{subfig}
\usepackage{amscd}

\newcommand{\beq}{\begin{equation}}
\newcommand{\eeq}{\end{equation}}
\newcommand{\ba}{\begin{array}}
\newcommand{\ea}{\end{array}}
\newcommand{\bea}{\begin{eqnarray}}
\newcommand{\eea}{\end{eqnarray}}

\begin{document}

\begin{center}

{\large \sc \bf {Simulation of multiple-quantum NMR dynamics of spin dimer   on 
  quantum computer}}

\vskip 15pt

{\large 
S.I.Doronin, E.B.Fel'dman, E.I.Kuznetsova, and  A.I.~Zenchuk
}

\vskip 8pt

{\it Institute of Problems of Chemical Physics, RAS,
Chernogolovka, Moscow reg., 142432, Russia}.

\vskip 8pt

\end{center}

\begin{abstract}
 Dymanics of spin dimers in multiple quantum NMR experiment is studied on the 5-qubit superconducting quantum processor of IBM {Quantum Experience} for the both  {pure} ground and thermodynamic equilibrium  (mixed) initial states. The work can be considered as a first step towards an application of quantum computers to solving problems of magnetic resonance. This article is dedicated to Prof. Klaus M\"obius and Prof. Kev Salikhov on
the occasion of their 85th birthdays.
\end{abstract}

{\bf Keywords:} spin dimer, multiple quantum NMR, quantum computer, pure and mixed states, thermodynamic equilibrium

\maketitle

\section{Introduction}
{ Quantum computers  based on quantum phenomena such as superposition and entanglement \cite{NCh} are expected to perform tasks which surpass the capabilities of modern classical digital computers \cite{Preskill}. First quantum computers arose quite recently and refer to noisy intermediate-scale quantum (NISQ) technology.} Quantum calculations open new possibilities for solving problems of magnetic resonance including dynamics of {many}-body systems \cite{BVFF}. Although the accuracy of today's calculations on quantum computers  is insufficient owing to errors of quantum gates it is {still} possible to perform quantum calculations for some relatively simple tasks \cite{CWSCGZLLP,BKRLDAW,ZSCetal,ZRPL,ZKEPL,DFZ_2020,BDFZ_2020}. Taking into account fantastic advantages of quantum computers over their classical counterparts which { is expected} to be released in future,  development of quantum algorithms is a challenging and useful task.  

In the present paper, we investigate { the { multiple-quantum (MQ) NMR} dynamics of a spin dimer} \cite{BMGP} on a quantum computer.  The { chosen} initial state of a spin dimer can be either the pure {ground} state  or the thermodynamic equilibrium (mixed) one. 

The article is organized as follows. In Sec.\ref{Section:principles},
 the principles of quantum algorithms and the main quantum gates are presented. The introduction to the theory of  {spin-dimers  dynamics}  in the MQ NMR experiment is given in Sec.\ref{Section:dynamics}
for different initial states of a dimer. {In Sec.\ref{Section:pure}, quantum calculation of the MQ NMR  dynamics of a spin dimer with the pure ground  and thermodynamic equilibrium initial states is given.} We briefly discuss our results in the concluding section \ref{Section:conclusion}.

\section{Main quantum gates and quantum algorithms }
\label{Section:principles}
The states of a qubit can be written as $|0\rangle$ or $|1\rangle$ { which  correspond to spin ($s=1/2$), respectively,  up and down}. Consequently, a two-qubit system considered below  has four computational basis states denoted as $|00\rangle$, $|01\rangle$, $|10\rangle$ and  $|11\rangle$. A pair of qubits can also exist in superposition of these four  states such that the state vector describing the two qubits is 
\begin{eqnarray}
|\psi\rangle =\alpha_{00} |00\rangle+ \alpha_{01} |01\rangle + \alpha_{10}|10\rangle+\alpha_{11}|11\rangle,\;\;
{|\alpha_{00}|^2 +|\alpha_{01}|^2 +|\alpha_{10}|^2   +|\alpha_{11}|^2=1,}
\end{eqnarray}
where the amplitudes $\alpha_{00}$, $\alpha_{01}$, $\alpha_{10}$ and $\alpha_{11}$  are the complex numbers. 

Different single qubit gates are considered in \cite{NCh}. We will use below one-qubit rotations and { two-qubit controlled NOT (CNOT) operation} \cite{NCh}. For example, the matrix representation of the rotation operator $R_x(\theta)$ by an angel $\theta$ about the axis $x$ can be written as 
\begin{eqnarray}\label{rx}
R_x(\theta) = \exp\Big(- i\frac{\theta}{2} \sigma_x\Big) =\left(
\begin{array}{cc}
\cos\frac{\theta}{2} & -i \sin\frac{\theta}{2}\cr
{-}i \sin\frac{\theta}{2}&\cos\frac{\theta}{2}
\end{array}
\right),
\end{eqnarray}
where $\sigma_x$ is the Pauli operator \cite{NCh}. The rotation operators $R_y$ and $R_z$ are defined analogously:
\begin{eqnarray}\label{ry}
R_y(\theta) = \exp\Big(- i\frac{\theta}{2} \sigma_y\Big) ,\;\; R_z(\theta) = \exp\Big(- i\frac{\theta}{2} \sigma_z\Big),
\end{eqnarray}
where $\sigma_y$, $\sigma_z$ are the corresponding Pauli operators.
The CNOT is very important in quantum computing. It can be used to entangle and disentangle different quantum states. Moreover, according to the Solovay - Kitaev theorem \cite{NCh} any quantum circuit can be simulated to an arbitrary degree of accuracy using a combination of CNOT gates and one-qubit rotations. Thus, one-qubit rotations and CNOT form the basis of an arbitrary quantum  algorithm. 

The CNOT gate operates on a quantum register consisting of two qubits.  {It} flips the second qubit (the target qubit) if and only if the state of the  first qubit (the control qubit) is $|1\rangle$. The action of the  CNOT gate can be represented in the computational basis by the matrix 
\begin{eqnarray}\label{cnot}
CNOT=\left(
\begin{array}{cccc}
1&0&0&0\cr
0&1&0&0\cr
0&0&0&1\cr
0&0&1&0
\end{array}
\right).
\end{eqnarray}
Below these gates are used for  investigation of dynamics of spin dimer in the MQ NMR experiment. 

\section{MQ NMR dynamics of spin dimers \cite{Doronin}   with a pure and thermodynamic equilibrium initial states}
\label{Section:dynamics}
MQ NMR dynamics of spin dimers in solids is described by either  the Shrodinger equation 
\begin{eqnarray}\label{L}
i\frac{d |\psi(t)\rangle }{dt}=H_{12} |\psi(t)\rangle
\end{eqnarray}
in the case of a pure initial state, or
the Liouville-von Neumann equation
\begin{eqnarray}\label{Sh}
i\frac{d \rho(t)}{dt}=[H_{12},\rho(t)]
\end{eqnarray}
in the case of a mixed initial state. Here $\rho(t)$ is the density matrix and 
the two-spin/two-quantum nonsecular average dipolar Hamiltonian $H_{12}$ is given by \cite{BMGP}
\begin{eqnarray}\label{H12}
H_{12}=-\frac{1}{2} D( I_1^+I_2^+ +  I_1^-I_2^-),
\end{eqnarray}
where $I_i^+$ and $I_i^-$, $i=1,2$, are raising and lowering  angular momentum operators  of dimer's spins and $D$ is the dipolar coupling constant.

{
\subsection{Pure ground state}}
 Let the spin system be in the pure ground state at $t=0$,
\begin{eqnarray}\label{inpure}
|\psi(0)\rangle = | 00\rangle.
\end{eqnarray}
Then the solution of Eq. (\ref{L}) can be written as 
\begin{eqnarray}\label{evH}
|\psi(t)\rangle= e^{-iH_{12} t} { |00\rangle.}
\end{eqnarray}
Calculation with (\ref{H12}), (\ref{inpure}), (\ref{evH})
leads to the simple result:
\begin{eqnarray}\label{psit}
|\psi(\tau)\rangle = \cos\frac{\tau}{2} |00\rangle + i\sin\frac{\tau}{2} |11\rangle,\;\;\tau=D t.
\end{eqnarray}
Presentation (\ref{psit})  is very useful for performing quantum calculation.

{
Now we find the intensities of MQ NMR coherences assotiated with  state (\ref{psit}). For this purpose we write the appropriate density matrix }
\begin{eqnarray}\label{rhot}
&&
\rho(\tau) = |\psi(\tau)\rangle \langle\psi(\tau)| =\\\nonumber
&&
 \cos^2\frac{\tau}{2}   |00\rangle \langle 00|     +  
  \sin^2\frac{\tau}{2}  |11\rangle \langle 11| 
{+}\frac{i}{2}  \sin\, \tau   |11\rangle \langle 00| {-}
\frac{i}{2}  \sin\, \tau   |00\rangle \langle 11|.
\end{eqnarray}
{By virtue of Eq. (\ref{rhot}) and taking into account that} the signal of the longitudinal magnetization is  observed in the MQ NMR experiment \cite{BMGP}, we find  the intensity $J_0(\tau)$  of the 0-order MQ NMR coherence {\cite{Doronin}},
\begin{eqnarray}\label{J01}
J_0= {\cos^2\frac{\tau}{2}} \cos \tau - {\sin^2\frac{\tau}{2}} \cos \tau = \cos^2\tau,
\end{eqnarray}
and the intensities $J_{\pm 2}(\tau)$ of  the $\pm 2$-order MQ NMR coherences,
\begin{eqnarray}\label{J21}
J_{\pm 2}=  \frac{\sin^2\tau}{2}.
\end{eqnarray}

{
\subsection{Thermodynamic equilibrium initial state}}
We consider now a spin dimer in a strong external magnetic field \cite{FP}. The thermodynamic equilibrium density matrix $\rho(0)$  of the system is
\begin{eqnarray}\label{equil}
\rho(0)=\frac{e^{\beta I_z}}{Z},\;\;Z={\mbox{Tr}} e^{\beta I_z}, \;\beta= \frac{\hbar \omega_0}{kT},
\end{eqnarray}
where $\omega_0$ is the Larmor frequency, $T$ is the temperature, $\hbar$ and $k$ are, respectively, the Plank and Boltzmann constants, {  $I_z=
 I_{1z } + I_{2z}$, $I_{jz}$ ($j=1,2$)  is the   projection of the angular  momentum operator of the spin $j$ on the axis $z$,} and
$Z$ is the partition function.  
The dimer density matrix $\rho(t)$ can be written in the computational basis as follows: 
\begin{eqnarray}\label{rhot2}
&&
\rho({\tau})= e^{-{\frac{i  H_{12}}{D}\tau}} \rho(0) e^{{\frac{i H_{12}}{D}\tau}} =\\\nonumber
&&
\frac{1}{2(1+\cosh\,\beta)}
\left(
\begin{array}{cccc}
\cosh\,\beta+ \cos\,\tau \sinh\,\beta&0&0& {-} i\sin\,\tau \sinh\,\beta\cr
0&1&0&0\cr
0&0&1&0\cr
i\sin\,\tau \sinh\,\beta&0&0&\cosh\,\beta- \cos\,\tau \sinh\,\beta
\end{array}
\right)
\end{eqnarray}
{ The diagonal part of the density matrix (\ref{rhot2})  is responsible for the intensity $J_0(\tau)$ of the 0-order coherence matrix, and the off-diagonal parts, $\rho_{\pm 2}=\frac{{\mp} i\sin\,\tau \sinh\,\beta}{{2(1+\cosh\,\beta)}}$, are responsible for the intensities   $J_{\pm 2}(\tau)$ of the $\pm 2$-order coherence matrices. } These  intensities are  following:
\begin{eqnarray}\label{JJ02}
&&
J_0= \cos^2 \tau \tan\frac{\beta}{2},\\\nonumber
&&
J_{{\pm 2}}= \frac{1}{2}\sin^2 \tau  \tan\frac{\beta}{2}.
\end{eqnarray}
{ Notice that  formulae (\ref{J01}) and (\ref{J21}) for the intensities $J_0$ and $J_2$ of the dimer with the pure ground initial state are  the low-temperature limits $\beta \to \infty$ of formulae (\ref{JJ02}).}

{ \section{Simulation of spin-dimer  dynamics on quantum computer}\label{Section:pure}
\label{Section:teq}
To simulate the evolution operators on a quantum processor, they must be represented in terms of one-qubit rotations  and CNOTs. We perform this simulation for both initial states considered above. { All calculations are performed on the 5-qubit quantum processor of IBM QE.}

\subsection{Pure ground  initial state}
}

\begin{figure*}[!]
\epsfig{file=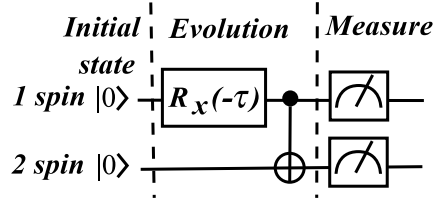,
  scale=0.4
   ,angle=0
} 
\caption{The scheme for simulating the dynamics  of the pure ground state of a spin-dimer  on a quantum processor } 
  \label{Fig:QE2} 
\end{figure*}

\begin{figure*}[!]
\epsfig{file=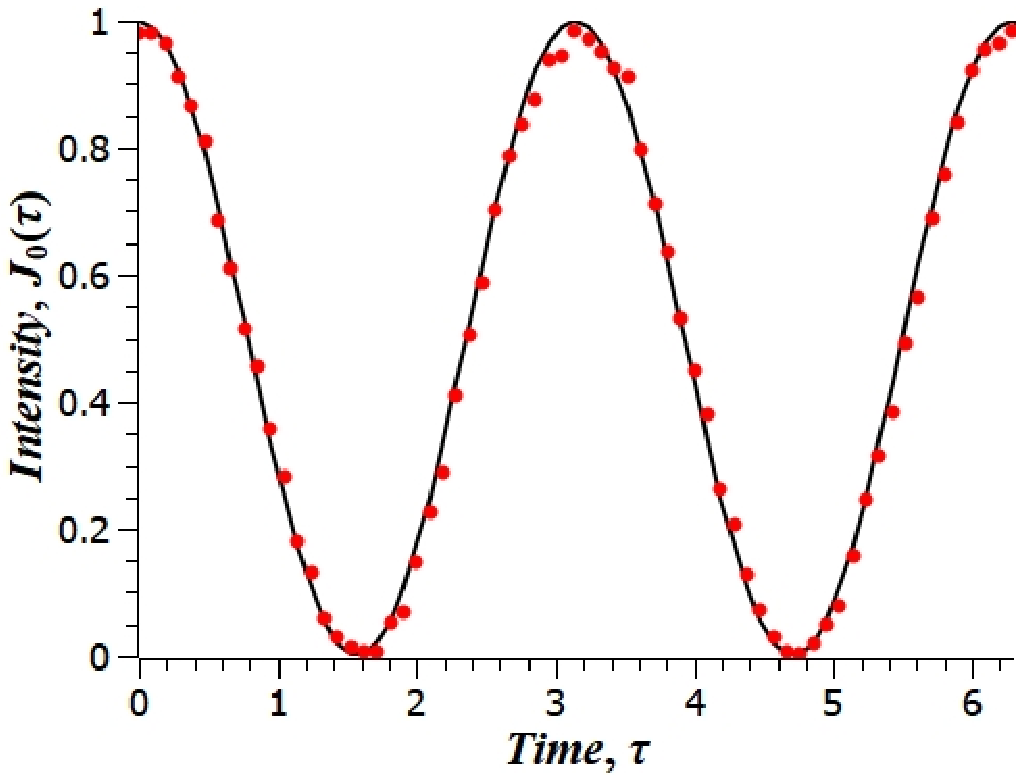,
  scale=0.6
   ,angle=0
} 
\epsfig{file=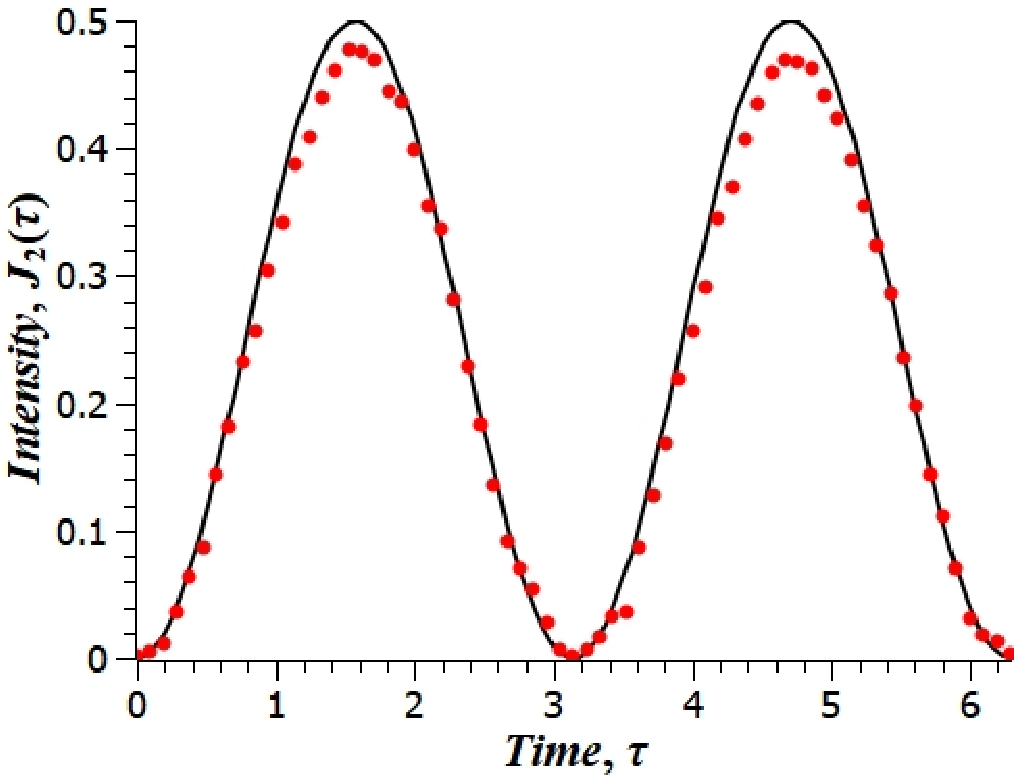,
  scale=0.6
   ,angle=0
} 
\caption{{Pure ground initial state.} The evolution of the intensities of the MQ NMR coherences of the 0- and  2-order ($J_0$ and $J_2$). Intensities obtained  theoretically using Eqs.(\ref{J01}) and (\ref{J21})  (solid line) slightly differ from the intensities found using results of calculation on the quantum processor  (circles), see Eqs.(\ref{J0J2a}). } 
  \label{Fig:J0J2} 
\end{figure*}

The evolution of a pure ground state (\ref{psit}) can be presented as follows:
\begin{eqnarray}\label{pc}
|\psi(\tau)\rangle =C_{12} R_{x1}({-}\tau)|00\rangle. 
\end{eqnarray}
This  evolution can be simulated on a quantum processor according to the scheme in   Fig.\ref{Fig:QE2}.
After the rotation of the first qubit by the angle $\theta={-\tau} $ about  the axes $x$ (see Eq.(\ref{rx})) the initial state changes as follows:
\begin{eqnarray}\label{rotation}
|00\rangle \to \cos{\frac{\tau}{2}} |00\rangle {+} i \sin{\frac{\tau}{2}} |10\rangle. 
\end{eqnarray}
After applying CNOT (\ref{cnot}) to state (\ref{rotation}) using the first spin as a control qubit, we obtain the following state of the system:
\begin{eqnarray}\label{thetapsi}
 \cos{\frac{\tau}{2}} |00\rangle {+} i \sin{\frac{\tau}{2}} |10\rangle  \to
 \cos{\frac{\tau}{2}} |00\rangle {+} i \sin{\frac{\tau}{2}}|11\rangle \equiv |\psi(\tau)\rangle . 
\end{eqnarray}
We see from (\ref{thetapsi}) that the probabilities of states $|00\rangle$ and $|11\rangle$ are 
\begin{eqnarray}\label{aa}
a_1(\tau) =\cos^2\frac{\tau}{2},\;\;
a_2(\tau) =\sin^2\frac{\tau}{2}.
\end{eqnarray}
Using Eqs.(\ref{J01}), (\ref{J21}) and (\ref{aa}) 
we obtain the intensities of 0- and ${\pm 2}$-order coherences {  in terms of the measured probabilities $a_i$, $i=1,2$:}
\begin{eqnarray}\label{J0J2a}
J_0(\tau) =(2 a_1(\tau) -1)^2,\;\;J_{{\pm 2}}(\tau) = 2 a_1(\tau) a_2(\tau).
\end{eqnarray}
In Fig.\ref{Fig:J0J2}, we compare the theoretical intensities of MQ NMR coherences of the 0- and 2-order (see Eqs.(\ref{J01}) and  (\ref{J21})) with the intensities (\ref{J0J2a}) found using results of calculation on the quantum processor. One can see a good agreement between {both} results.

\begin{figure*}[!]
\epsfig{file=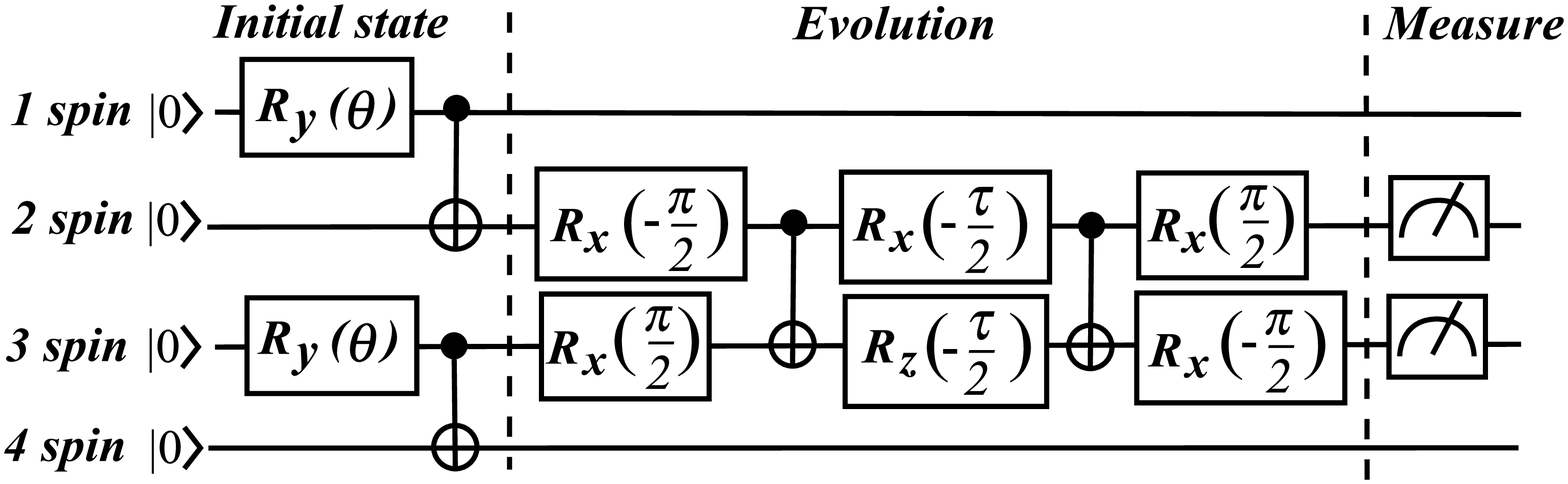,
  scale=0.4
   ,angle=0
} 
\caption{The scheme for simulating the  dynamics of the dimer's  thermodynamic equilibrium  initial state  on a quantum processor} 
  \label{Fig:QE1} 
\end{figure*}

\begin{figure*}[!]
\epsfig{file=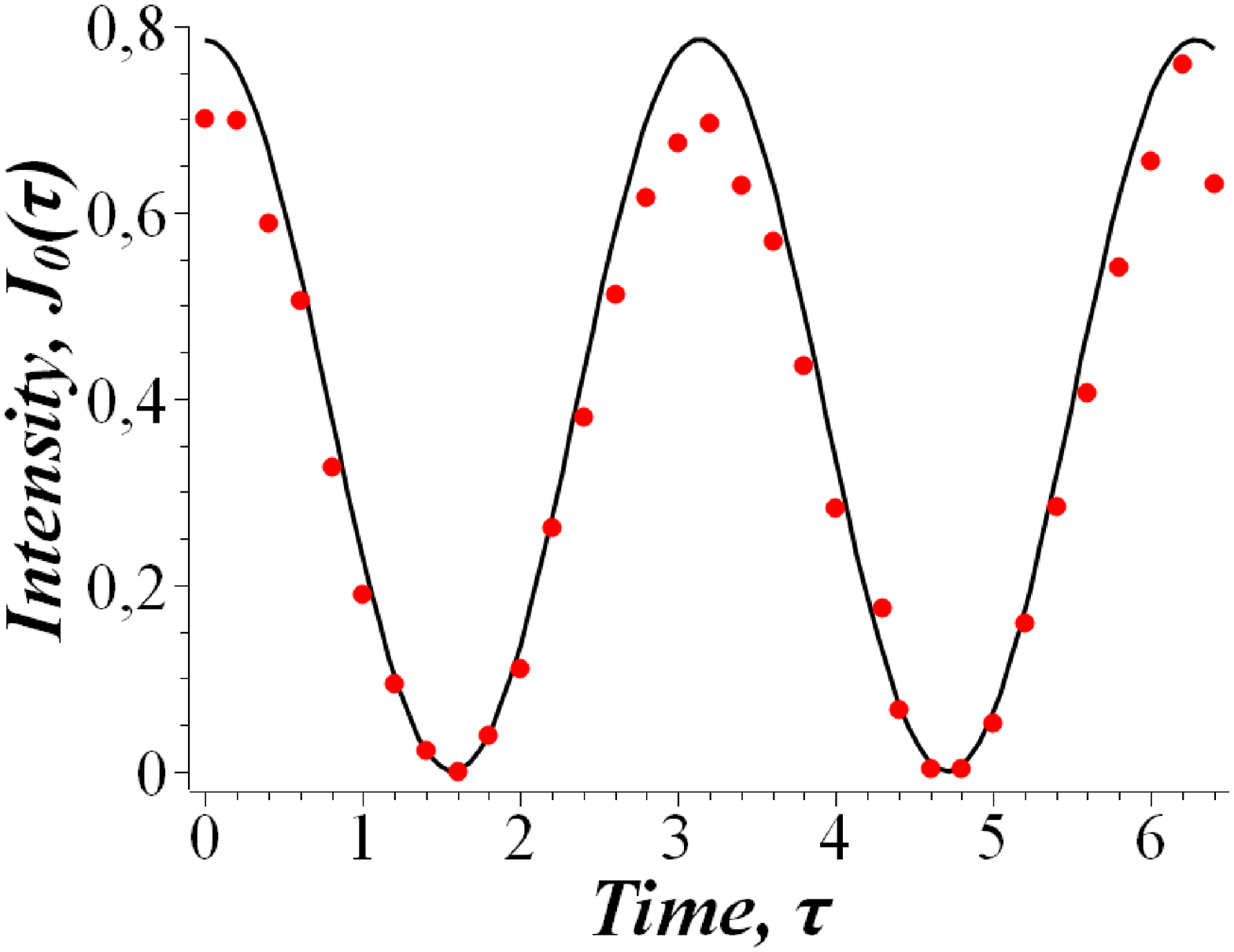,
  scale=0.3
   ,angle=0
}
\epsfig{file=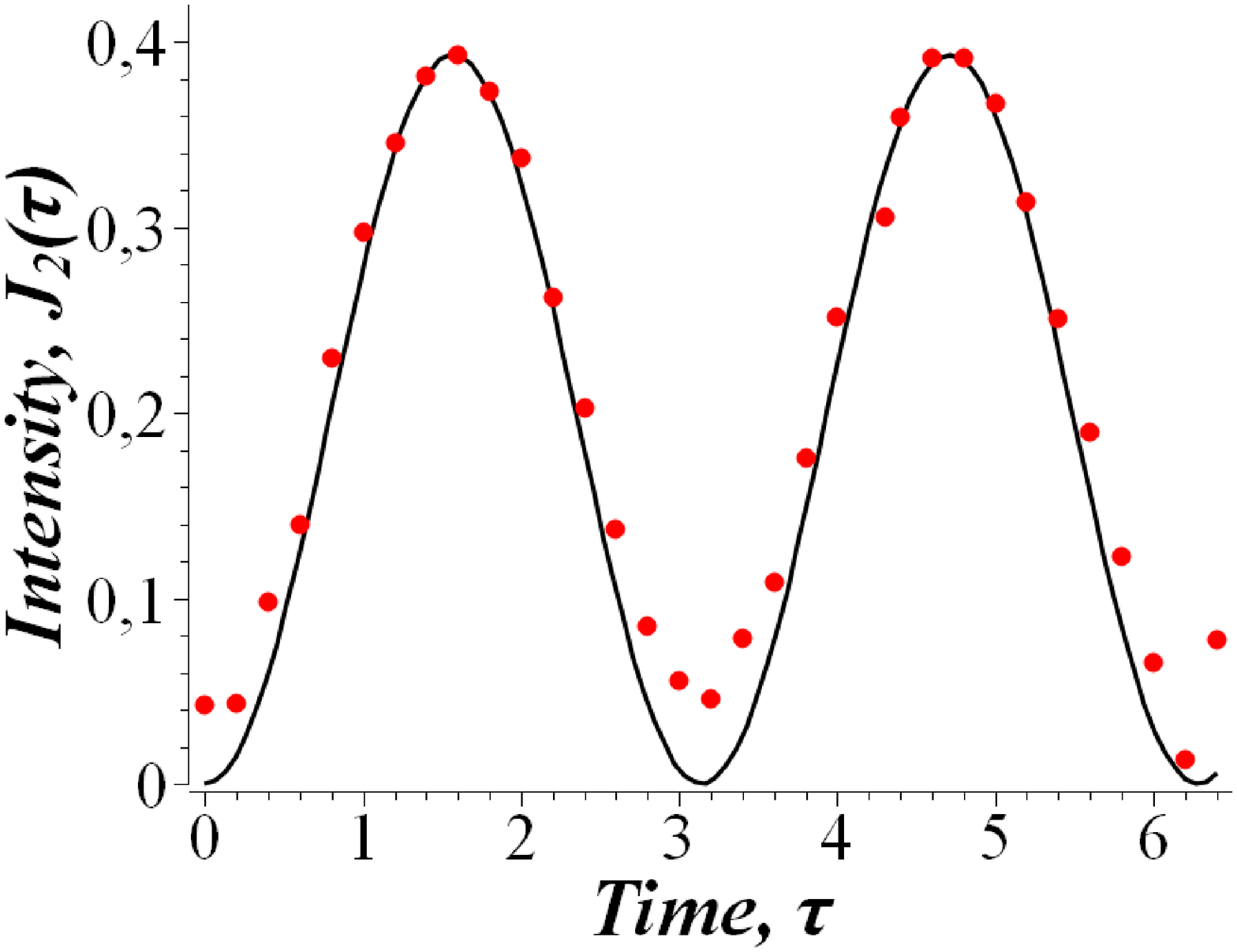,
  scale=0.3
   ,angle=0
} 
\caption{{Thermodynamic equilibrium initial state with {$\beta=2.12$}.}
The evolution of the intensities of the MQ NMR coherences of the 0- and  2-order ($J_0$ and $J_2$). Intensities obtained  theoretically using Eqs.(\ref{JJ02})  (solid line) slightly differ from the intensities found using results of calculation on the quantum processor  (circles), see Eqs.(\ref{J0b}) and (\ref{J2b}).
} 
  \label{Fig:teq} 
\end{figure*}

{
\subsection{Thermodynamic equilibrium initial state}
}
The dimer's thermodynamic equilibrium initial state (\ref{equil})  can be considered as a tensor product of 1-qubit states:
\begin{eqnarray}\label{rho0}
&&\rho(0)= \rho_1(0) \otimes  \rho_{2}(0) , \;\;\\\label{rho01}
&&\rho_k(0) = \frac{e^{\beta I_{k z}}}{Z_k},\;\;Z_k={\mbox{Tr}} e^{\beta I_{kz}},\;\;{k=1,2}.
\end{eqnarray}
{However, we can simulate the evolution only of a pure state on a quantum processor.  Therefore, we have to purify the initial state (\ref{rho0}) \cite{NCh}. It is simple to show that one additional spin is enough to purify the thermodynamic equilibrium state of a single spin. 
In fact, let  $|\psi_{12}\rangle$ be  the 2-qubit  state of the form
\begin{eqnarray}\label{psi1234}
|\psi_{12}\rangle=\cos\frac{\theta}{2} |00\rangle+ \sin\frac{\theta}{2} |11\rangle.
\end{eqnarray}
Tracing  the density matrix $|\psi_{12}\rangle \langle \psi_{12}|$  over one of  qubits yields the following  1-qubit state:
\begin{eqnarray}
{\mbox{diag}}(\cos^2\frac{\theta}{2}, \sin^2\frac{\theta}{2})
\end{eqnarray}
which coincides with  $\rho_k(0)$ (\ref{rho01}), $k=1,2$, if
\begin{eqnarray}\label{cos}
\cos^2\frac{\theta}{2} = \frac{e^{{\beta/2}}}{2\cosh\,\beta/2}\;\;\Leftrightarrow \;\;
\cos\,\theta = \tanh\frac{\beta}{2}.
\end{eqnarray}
Therefore, the  purification of  2-qubit state (\ref{rho0}) yields  the following pure state of the 4-qubit system:}
\begin{eqnarray}
|\psi(0)\rangle = (\cos\frac{\theta}{2} |00\rangle+ \sin\frac{\theta}{2} |11\rangle)\otimes
(\cos\frac{\theta}{2} |00\rangle+ \sin\frac{\theta}{2} |11\rangle),
\end{eqnarray}
which can be prepared on a quantum processor  as follows:
\begin{eqnarray}\label{QE1}
|\psi(0)\rangle =  C_{12} R_{y1}(\theta) C_{34} R_{y3}(\theta) |0000\rangle,
\end{eqnarray}
where $C_{ij}$, $j>i$ is the  CNOT  (\ref{cnot}) and  $R_{ky}(\theta)$ ($k=1,3$) is the rotation operator by an angle $\theta$ about the axis $y$. 
We consider the thermodynamic equilibrium state of the dimer consisting of the 2nd and 3rd qubits {of the 5-qubit register} assuming tracing over the 1st and 4th spins {(the 5th qubit is not included into the scheme)}. 
Therefore we apply the evolution operator to the 2nd and 3rd spins only:
\begin{eqnarray}
|\psi(t)\rangle = e^{-iH_{23} t} |\psi(0)\rangle.
\end{eqnarray}
By virtue of the  result of Ref.\cite{VD}, we can write
\begin{eqnarray}\label{QE2}
e^{-iH_{23} t} = R_{x2}(\frac{\pi}{2}) R_{x3}(-\frac{\pi}{2}) C_{23} R_{x2}(-\tau/2) R_{z3}(-\tau/2) C_{23} R_{x2}(-\frac{\pi}{2}) R_{x3}(\frac{\pi}{2}).
\end{eqnarray}
{Using the scheme in Fig.\ref{Fig:QE1}, we simulate   the initial state (\ref{QE1}) and evolution (\ref{QE2}) on a quantum computer.}
As a result, we obtain the following pure state of four qubits
\begin{eqnarray}\label{psi1234_2}
|\psi(t)\rangle &=& \cos\frac{\tau}{2} \cos^2 \frac{\theta}{2} |0000\rangle +
\frac{1}{2} \sin \theta |0011\rangle + i \sin\frac{\tau}{2} \cos^2 \frac{\theta}{2} |0110\rangle+\\\nonumber
&&
i \sin\frac{\tau}{2} \sin^2 \frac{\theta}{2} |1001\rangle+
\frac{1}{2} \sin \theta |1100\rangle +\cos\frac{\tau}{2} \sin^2 \frac{\theta}{2} |1111\rangle.
\end{eqnarray}
Measurements over the 2nd and 3rd spins yield the probabilities $p_{nm}$ of their states:
{
\begin{eqnarray}\label{pnm}
p_{nm}=\sum_{k,l=0,1}|\langle k nm l|\psi(\tau)\rangle|^2,\;\;n,m=0,1,
\end{eqnarray}
}
where the sum is over the states of the first and fourth spins.
By virtue of Eq.(\ref{psi1234_2}), Eq.(\ref{pnm}) yields:
\begin{eqnarray}
&&
p_{00}=\frac{1}{8}(3+4\cos\,\tau \cos\,\theta + \cos(2\theta)),\\\nonumber
&&
p_{11}=\frac{1}{8}(3-4\cos\,\tau \cos\,\theta + \cos(2\theta)),\\\nonumber
&&
p_{01}=p_{10}=\frac{\sin^2\theta}{4},
\end{eqnarray}
which are the diagonal elements of the 2-qubit  reduced density matrix of the  2nd and 3rd spins:
\begin{eqnarray}
\rho_{23}&=&Tr_{1,4}|\psi(\tau)\rangle\langle\psi(\tau)|=\\\nonumber
&&
\left(
\begin{array}{cccc}
p_{00}&0&0&\rho_{14}\cr
0&p_{01}&0&0\cr
0&0&p_{10}&0\cr
\rho_{14}^*&0&0&p_{11}
\end{array}
\right),\;\;\rho_{14}=-\frac{i}{2}\sin \tau \cos \theta.
\end{eqnarray}
For the intensity $J_0$ of the  0-order MQ NMR coherence,  according to the standard rules \cite{FP} and taking into account (\ref{cos}), we obtain
\begin{eqnarray}\label{J0b}
&&
J_0(\tau)={ \cos\tau}  (p_{00}-p_{11}).
\end{eqnarray}
The conservation low of the sum of MQ NMR coherences  \cite{FP} allows to calculate the intensities of the ${\pm}2$-order MQ NMR coherence:
\begin{eqnarray}\label{J2b}
&&
J_{\pm 2}(\tau)= \frac{1}{2}\Big(  \tan\frac{\beta}{2}-J_0(\tau)\Big).
\end{eqnarray}
We compare the  intensities {(\ref{J0b}) and (\ref{J2b})} found using the results of calculation on the quantum computer with the theoretical results (\ref{JJ02}). One can see a good agreement between  {them} in Fig.\ref{Fig:teq}.

\section{Conclusion}
\label{Section:conclusion}

{We investigate the spin-dimer dynamics on a 5-qubit platform of IBM superconducting quantum computer. Two initial states are considered: the pure ground state and the thermodynamic equilibrium one.  Intensities of the 0- and 2-order MQ NMR coherences found on the basis of calculation on the quantum computer are compared with theoretically obtained intensities and   a good agreement between them is  demonstrated. }

We believe that quantum computer calculations open large perspectives in solving  problems of spin dynamics and magnetic resonance.

{\bf Acknowledgments} { Authors acknowledge the use of the IBM Quantum Experience for this work. The viewpoints
expressed are those of the authors and do not reflect the official policy or position of IBM.}
We acknowledge funding from the Ministry of Science and Higher Education  of the Russian Federation (Grant No. 075-15-2020-779).

\end{document}